\documentclass[epj]{svjour}
\usepackage{latexsym}
\usepackage{graphics}

\begin{document}

\title{Magnetic Field Tomography}
\author{Ph.W. Courteille\inst{1} \and S.R. Muniz\inst{1} \and K. Magalh\~{a}es\inst{1} \and R. Kaiser\inst{2} \and
L.G. Marcassa\inst{1}, and V.S. Bagnato\inst{1}}

\institute{Instituto de F\'{i}sica de S\~{a}o Carlos, USP\\
Caixa Postal 369, CEP 13560-970, S\~{a}o Carlos/SP, Brazil\\
Phone 55-16-2712012: Fax 55-16-2713616,
\email{Philippe.Courteille@up.univ-mrs.fr}\\
\and
Laboratoire Ondes et D\'{e}sordre FRE 2302 CNRS\\
1361, Route des Lucioles, F-06560 Valbonne}

\date{Received: date / Revised version: date}

\abstract{ Neutral atoms may be trapped via the interaction of
their magnetic dipole moment with magnetic field gradients. One
of the possible schemes is the cloverleaf trap. It is often
desirable to have at hand a fast and precise technique for
measuring the magnetic field distribution. We use for
instantaneous imaging the equipotential lines of the magnetic
field a diagnostic tool which is based on spatially resolved
observation of the fluorescence emitted by a hot beam of sodium
atoms crossing a thin slice of resonant laser light within the
magnetic field region to be investigated. The inhomogeneous
magnetic field spatially modulates the resonance condition
between the Zeeman-shifted hyperfine sublevels and the laser
light and therefore the amount of scattered photons. We apply
this technique for mapping the field of our cloverleaf trap in
three dimensions under various conditions.}


\maketitle{}

\section{Introduction}
\label{intro}

For two decades, the invention of cooling and trapping techniques
for neutral atoms boosts the field of cold atomic physics and
several years ago
even led to the realization of the long expected goal of \textit{%
Bose-Einstein condensation} (BEC) in dilute atomic gases,
followed by a wealth of remarkable experiments\ on atom lasing
and superfluidity of condensed gases. Most trapping
configurations are based on light forces or magnetic field
gradients or combinations of both. The most frequently used trap
for neutral atom trapping is the \textit{Magneto-Optical Trap}
(MOT) \cite{Raab87}. MOTs have allowed the confinement of more
than $10^{10}$ atoms \cite{Gibble92} at temperatures well below
$100~\mathrm{\mu K}$ and densities up to
$10^{12}~\mathrm{cm}^{-3}$. Radiation trapping by the optically
thick cloud, however, sets a limit to further compression due to
photon rescattering \cite{Walker90}. Even employing more
sophisticated schemes \cite{Ido00}, the highest phase space
density achieved today remains at about one order of magnitude
below the threshold to Bose-Einstein condensation. The phenomenon
of radiation trapping is intrinsically connected to the fact that
in magneto-optical traps the restoring force comes from the
radiation pressure exerted by the laser beams. Alternative
trapping schemes based on the dipole force of far-detuned laser
beams avoid the radiation trapping problem \cite{Chu86,Han00}. A
different approach is to exploit the weak interaction of the
dipole moment of paramagnetic atoms with magnetic field
gradients. Magnetic fields can be designed to generate local
minima in free space capable of acting as trapping potentials.
Several types of magnetic traps have been successfully used in
BEC experiments like the \textit{Time-Orbiting Potential} (TOP)
trap \cite{Petrich95} and various \textit{Ioffe-Pritchard} (IP)
type traps \cite{Pritchard83}, \cite{Mewes96}. Because magnetic
traps are generally weak, large magnetic fields and loading with
precooled atoms is required. Generally the atoms are precooled in
a standard MOT and then transferred into the magnetic trap. In
the absence of laser light the magnetic trap requires a new kind
of cooling mechanism. The only mechanism that has successfully
led to BEC up to now is \textit{Forced Radiofrequency
Evaporation} \cite{Hess86,Anderson95}. A necessary condition for
evaporation to work is that the rate of elastic collisions
between the trapped atoms be large enough to maintain the cloud
in thermal equilibrium during the evaporation process which must
take place within a period shorter than the trap lifetime
\cite{Monroe93}. Large collision rates mean big atomic clouds
and/or steep trapping potentials.

This work is motivated by our need to characterize our
\textit{cloverleaf trap} which is, in principle, a
Ioffe-Pritchard type trap \cite{Ketterle99}. The necessity of
optimizing the efficiency for loading, trapping and evaporative
cooling sets tight conditions for the design of the magnetic
coils and the current supplying circuitry. \emph{E.g.}, the field
gradient must be rather strong and capable of being varied over a
wide range. It is important to calibrate the secular frequencies,
which essentially govern the shape of the trapped atomic cloud,
as a function of the control parameters. Therefore we need a
precise control of the fields and a suitable way of measuring and
monitoring them. Measuring magnetic fields is usually performed
with Hall probes. However, very often the region of space to be
analyzed is within a vacuum recipient and thus not accessible to
massive probes. We use the idea of fluorescence imaging magnetic
fields \cite{Kaenders96} to develop a simple technique for the
instantaneous mapping of magnetic field equipotential lines in
two dimensions within a closed vacuum recipient. This technique
that we call \textit{Magnetic Field Tomography} (MFT) can even be
extended to allow a three-dimensional reconstruction of the
complete magnetic vector field. The method provides a simple and
reliable diagnostic tool for quantifying the field distribution
of our magnetic trap for any set of values of the control
parameters.

This paper is organized as follows. In section~2, we present our
magnetic trapping experiment and briefly expose our approach to
calculating the magnetic field distribution. In section~3, we
introduce our Magnetic Field Tomography technique. We tested the
technique and present experimental results in section~4, where we
also show how to process the experimental data and compare the
results to the calculated fields. We conclude our paper with a
general discussion.

\section{The Magnetic Trap}
\label{sec:1}

Magnetic traps are usually built with current carrying coils or
permanent magnets \cite{Tollett95}. In the first case, high
magnetic field gradients generally require high currents flowing
through the coils. In special cases, high gradients may be
achieved with lower currents \cite{Desruelle98,Haensch99}. The
main constraints for the design of the coils for the cloverleaf
trap arise from the geometry of the vacuum chamber and by the
power supply available: in our setup, the minimum distance
between the coils and the center of the trap is $15~\mathrm{mm}$.
We apply up to $300~\mathrm{A}$ and need to dissipate the
$10~\mathrm{kW}$ resistive heat by cooling the coils with a
high-pressure water flow passing through the hollow wires. Our
magnetic cloverleaf trap is schematically shown in Figure~1.
Radial confinement of the atoms is assured by four pairs of coils
in anti-Helmholtz configuration (off-center coils in Fig.~1)
producing a quadrupolar waveguide field along the symmetry axis.
Axial confinement is realized via a magnetic bottle (inner and
outer axially centered coils in Fig.~1). The (inner) pinch coils
create the necessary field curvature, and a pair of (outer)
antibias coils in Helmholtz configuration compensates the large
central offset produced by the pinch coils to a variable amount.
As said earlier, to initiate run-away evaporative cooling, we
must be able to realize large interatomic elastic collision
rates. This is achieved by compressing the magnetic trap: at the
time of loading the trap from the MOT the mean secular frequency
is typically $\omega _{trap}\approx 2\pi \times 20~\mathrm{Hz}$
(the transfer efficiency is best when the curvatures of the MOT
and the magnetic trap are approximately matched), and the secular
frequency of the fully compressed trap is $\omega _{trap}\approx
2\pi \times 200~\mathrm{Hz}$. In the case of the cloverleaf trap,
the compression is controlled via the antibias field.
\begin{figure} \vspace{6cm} \caption{Scheme of the
cloverleaf trap. The cloverleaf coils (blue) produce a radially
linear magnetic field. The pinch coils (red) produce an axially
confining magnetic bottle. The antibias coils (green) are
Helmholtz coils and serve to compensate the magnetic field
offset.} \label{fig:1}
\end{figure}

Precise control of the magnetic field bias is also important
because it fixes an offset for the resonant radiofrequency used
for evaporative cooling: The evaporating surface
\cite{Ketterle96} is set by the condition that the radiofrequency
balances the shift of the atomic energy levels
induced by the local magnetic field $\hbar \omega _{rf}=\left| \mathbf{\mu B}%
(\mathbf{r})\right| $. In practice, the magnetic trap is
compressed until the bias amounts to only a few Gauss. This means
that fluctuations should be very small which is not trivial
because the bias results from the subtraction of two large
magnetic fields, the magnetic bottle field and the antibias
field. For our setup this calls for a current stability superior
to $10^{-4}$. Furthermore, when loading the magnetic trap or
releasing the atoms, we need to be able to quickly switch on and
off the magnetic field, \emph{i.e.}$\ $typically faster than
within $100~\mathrm{\mu s}$ \cite {Anderson95}. In summary, the
current control circuitry has to fulfill several demanding
requirements, since we ask for high current, low noise,
independent tuning of part of the current, and fast switching.
The details of the current switch will be detailed elsewhere
\cite{Muniz01}.

Once the geometry of the coils is known, calculating the magnetic fields $%
\mathbf{B}(\mathbf{r})$ is very easy. For simple geometrical
shapes of the coils, one can use analytic formulae
\cite{Bergeman87} containing elliptical integrals. For
complicated shapes of the coils or asymmetric arrangements, we
numerically integrate the Biot-Savart formula
\begin{equation}
\mathbf{B}(\mathbf{r})=\frac{\mu _{0}I}{4\pi }\oint_{C}\frac{d\mathbf{s}%
\times \left( \mathbf{r}-\mathbf{s}\right) }{\left| \mathbf{r}-\mathbf{s}%
\right| ^{3}},  \label{EqBiotSavart}
\end{equation}
where $\mathbf{s=s}(\varsigma )$ is the current path parametrized
in small steps $\varsigma $. From such calculations, we determine
to first order the following data characterizing our cloverleaf
trap close to its center. The magnetic field gradient per ampere
produced by the current $I_{clov}$ flowing through the cloverleaf
coils is $\partial _{r}B_{clov}/I_{clov}=0.438~\mathrm{G/cm/A}$.
The curvature of the magnetic bottle field generated by the
current $I_{pinch}$ in the pinch coils is in
radial direction $\partial _{r}^{2}B_{pinch}/I_{pinch}=-0.186~\mathrm{G/cm}%
^{2}\mathrm{/A}$ and in axial direction $\partial
_{z}^{2}B_{pinch}/I_{pinch}=0.373~\mathrm{G/cm}^{2}\mathrm{/A}$.\
And the magnetic field amplitude produced by the pinch and the
antibias fields at
the trap center are $B_{pinch}/I_{pinch}=0.767~\mathrm{G/A}$ and $%
B_{anti}/I_{anti}=2.693~\mathrm{G/A}$. We usually work with the currents $%
I_{clov}=I_{pinch}=285~\mathrm{A}$ and $I_{anti}$\ variable.
Inaccuracies in the geometric shape of the coils limit the level
of precision of the calculations to $10\%$ uncertainty. However,
as we will show later, we can calibrate the calculations using
the MFT method and achieve uncertainties lower than $2\%$.

The absolute value of the magnetic fields
$|\mathbf{B}(\mathbf{r})|$ is shown in Figure~2 for two different
values of the antibias current. For low antibias current (left
side), $I_{anti}=50~\mathrm{A}$, the trap is roughly isotropic at
the center, the mean curvature of the magnetic field is small,
and there is a large magnetic field offset,
$B(0)=100~\mathrm{G}$. However, when we apply a high antibias
current, $I_{anti}=85~\mathrm{A}$, the trap is axially elongated,
radially compressed, and the magnetic field offset is small,
$B(0)=1~\mathrm{G}$.
\begin{figure} \vspace{6.5cm} \caption{Magnetic
trapping potentials in the isotropic and the compressed case. The
experimental parameters only differ in the amount of
antibias current: $I_{anti}=50~\mathrm{A}$ in the isotropic case and $%
I_{anti}=85~\mathrm{A}$ in the compressed case. The planes
slicing the potentials represent the laser being tuned to a
Zeeman shifted resonance. Our MFT technique images the contour
lines are at these detunings (see text).} \label{fig:2}
\end{figure}

\section{Magnetic Field Tomography}
\label{sec:2}

The basic idea of magnetic field tomography can be subsumed as
"laser excitation spectroscopy on Zeeman-shifted electronic
transitions with spatially resolved fluorescence detection". We
are detailing this idea in the following. The Hamiltonian of
atoms with the electron total angular
momentum $\mathbf{\mu }_{J}=g_{J}\mu _{B}\mathbf{J}$ and the nuclear spin $%
\mathbf{\mu }_{I}=g_{I}\mu _{N}\mathbf{I}$ subject to a magnetic field $%
\mathbf{B}$ reads \cite{Metcalf99}
\begin{eqnarray}
&&H_{hfs}+H_{B}  \label{EqIntermediateShift} \\
&=&A_{J}\mathbf{IJ}+B_{J}\frac{6(\mathbf{IJ})^{2}+3\mathbf{IJ}-2I(I+1)J(J+1)%
}{2I(2I-1)2J(2J-1)}  \nonumber \\
&&-\mathbf{\mu }_{J}\mathbf{B}-\mathbf{\mu }_{I}\mathbf{B}
\nonumber
\end{eqnarray}
The first term corresponds to the magnetic dipole interaction,
and $A_{J}$
is called the \textit{hyperfine constant}. For sodium it takes the values $%
A_{J}(3~^{2}S_{1/2})=885.8~\mathrm{MHz}$ and $A_{J}(3~^{2}P_{3/2})=18.65~%
\mathrm{MHz}$ \cite{Metcalf99}. The second term corresponds to the
electrostatic interaction of the nuclear quadrupole moment. The \textit{%
constant of the quadrupole interaction} for sodium, $%
B_{J}(3~^{2}P_{3/2})=2.82~\mathrm{MHz}$, is small but not
negligible. The next two terms describe the interaction of the
atom with the external magnetic field. The contribution
containing the nuclear magneton $\mu _{N}$ is negligibly small.
It is interesting to point out the behavior of the \textit{fully
stretched} spin states $|F=J+I,m_{F}=\pm F\rangle $. For those
states the shift due to the
interaction~(\ref{EqIntermediateShift}) is linear in
$|\mathbf{B}|$. The shifts of all other states, since our
magnetic trap operates in an intermediate regime of magnetic
fields where the hyperfine coupling is perturbed by the magnetic
fields, are conveniently numerically calculated as the eigenvalues
of the Hamiltonian. In order to keep the analysis simple, we will
first concentrate on the fully stretched states. However, we will
need to stress the full expression to describe several features
of our observations.

When a laser is irradiated on an electronic transition with Zeeman
degeneracy, the situation gets more complicated, because the
Zeeman substates can be coupled by the laser, optical pumping
takes place, and the energy levels are additionally shifted by
the dynamical Stark-effect by an amount that depends on the Rabi
frequency $\Omega =\mathbf{dE}/\hbar $, where $\mathbf{d}$ denotes
the electric dipole moment of the transition, and $%
\mathbf{E}$ denotes the electrical field amplitude of the light.
We will focus here on an approximation that holds in our
experiment, \emph{i.e.} the Rabi frequency is weak compared to the
Larmor frequency $\omega _{L}=\mathbf{\mu B}/\hbar $. Then the
energy levels are mainly shifted by the Zeeman-effect, and we may
disregard the light shift. In this case, the magnetic field
direction lends itself naturally as the quantization axis,
because the Hamiltonian is diagonal with respect to this axis.
\begin{figure}
\vspace{3cm} \caption{Level scheme of the sodium $D2$-line
between the hyperfine levels $F=2$ and $F^{\prime }=3$. The axial
magnetic field $B_{z}$ is assumed linear in the coordinate $z$.}
\label{fig:3}
\end{figure}

For a laser irradiated on an electronic resonance to induce
transitions, these transitions must be allowed by selection rules.
Electric dipole transitions \emph{e.g.} require $F^{\prime
}-F=0,\pm 1$ and $m_{F^{^{\prime }}}-m_{F}=0,\pm 1$, where the
prime denotes the excited level. In an external magnetic field
these transitions are not degenerate, so that magnetic field
probing via observation of the fluorescence as a function of
laser detuning is not unambiguous. If we choose our atomic
transition so that the total angular momentum of the upper level
is superior by $1$ to that of the lower level, $F^{\prime }=F+1$,
and if we apply a particular laser polarization, \emph{i.e.}
$\sigma ^{+}$ or $\sigma
^{-}$ respectively, we drive a so-called \textit{cycling transition}, \emph{%
i.e.} the excited state can only decay into the very same ground
state (see Fig.~3). The cycling transition involves the levels $S_{1/2},F=2,m_{F}=\pm 2$ and $%
P_{3/2},F^{\prime }=3,m_{F^{\prime }}=\pm 3$. Assume for example
a laser beam irradiated along the $z$%
-axis, $\mathbf{k}=k\mathbf{\hat{e}}_{z}$, inside a magnetic
field oriented in the same direction,
$\mathbf{B}=B_{z}\mathbf{\hat{e}}_{z}$. The choice of the
$z$-axis defines the labelling of the light polarization,
\emph{i.e.} the light polarization called $\sigma ^{-}$ is the
one that excites transitions with $m_{F^{\prime }}-m_{F}=-1$. Let
us consider a $\sigma ^{-}$ polarized red-detuned laser $\Delta
<0$. For atoms subject to a positive
magnetic field, $B_{z}>0$ the laser then drives the cycling transition $%
m_{F}=F=2$ and $m_{F^{\prime }}=F^{\prime }=3$, while atoms that
are subject to negative $B_{z}<0$ are off resonance. By inverting
either the polarization or the detuning, we probe the atoms that
are subject to negative $B_{z}$. This simple picture only
holds if we can assume to have a strong magnetic offset field in $z$%
-direction, $|\mathbf{B}|\approx |B_{z}|$, which is the case for
the field to be investigated here. We will see that the condition
is less restrictive than it seems in the first place.

Because our cycling transition effectively realizes a \textit{two-level atom}%
, the expression for the resonance fluorescence in terms of
scattered photon numbers $S$ per probing time $t_{probe}$ takes a
very simple form:
\begin{equation}
S=\frac{I}{\hbar \omega }\sigma N_{at}t_{probe},
\label{EqResonanceFluorescence}
\end{equation}
where $I$ is the intensity of the irradiated light, $N_{at}$ the
number of atoms in the interaction zone, and the optical cross
section is
\begin{equation}
\sigma =\frac{\sigma _{0}\Gamma ^{2}}{4\left[ \Delta -\hbar
^{-1}(\mathbf{\mu }^{\prime }-\mathbf{\mu })\mathbf{B}\right]
^{2}+2\Omega ^{2}+\Gamma ^{2}}, \label{EqCrossSection}
\end{equation}
where $\Gamma =2\pi \times 9.89~%
\mathrm{MHz}$ is the natural linewidth of the sodium $D2$ line, and $\Omega =%
\sqrt{\sigma _{0}\Gamma I/\hbar \omega }$ is the Rabi frequency.
For the cycling transition between the fully stretched states we
find $(\mathbf{\mu }^{\prime }-\mathbf{\mu })\mathbf{B}=\mu
_{B}(g_{F^{\prime }}m_{F^{\prime }}-g_{F}m_{F})\varsigma|\mathbf{B}|=\pm \mu _{B}\varsigma|%
\mathbf{B}|$ for $\sigma ^{\pm }$ polarized light, where
$\varsigma =\mathrm{sign}~(B_{z})$ and the Land\'{e} factors are $%
g_{F}=1/2$ and $g_{F^{\prime }}=2/3$.

The light is scattered at a nondegenerate transition between
magnetic sublevels so that we have to weight the coupling strength
(\emph{i.e.} the Rabi frequency) with the relative strength of
the specific transition. The relative coupling strength can be
expressed by $\{6j\}$ symbols which reflect the fine and the
hyperfine structure coupling and by $(3j)$\ symbols for the
coupling of the atomic angular momenta to the magnetic field
\cite {Metcalf99}. For the cycling transition of interest, the
relative coupling strength is $1$. In those cases where
$\mathbf{B}$ is not perfectly parallel to $\mathbf{\hat{e}_z}$,
the coupling strength depends on the relative orientation of the
polarization of the irradiated light and the magnetic field
direction $\mathbf{\hat{\varepsilon}\hat{B}}$. Those effects can
be included in our description by using a modified Rabi frequency $\Omega _{%
\mathbf{\hat{\varepsilon}\hat{B}}}$. However, we will see later
that we do not need to specify the functional dependence for our
purpose.
\begin{figure}
\vspace{4cm} \caption{Scheme of the tomographic measurement
experiment with atomic beam, laser light sheet and imaging
system. The symmetry axis of the cloverleaf trap is along
$\mathbf{\hat{e}}_z$. The imaging system is inclined by $22.5{
{{}^\circ}%
}$\ with respect to the atomic beam axis.}
\label{fig:4}
\end{figure}

The above considerations hold for single atoms interacting with a
magnetic and an optical field. In the following, we will address
the question how these atoms can be utilized for magnetic field
tomography. We consider
spatially inhomogeneous fields $\mathbf{B}(\mathbf{r})$ and $\mathbf{E}(%
\mathbf{r})$ and specify the physical situation further by
assuming that the spatial extend of the magnetic field is much
larger than the interaction zone of the atoms with the light
field. In fact, we will use a hot atomic beam perpendicularly
intersected by a thin sheet of laser light. This allows us to
disregard the impact of the magnetic field gradient on the atomic
motion. We define our coordinate system by calling the
propagation direction
of the laser beam $z$-axis and the normal vector of the light sheet plane $y$%
-axis (see Fig.~4). The light sheet may be displaced along the
$y$-axis by an amount $y_{0}$ and has an inhomogeneous Gaussian
intensity distribution along the $x$-axis so that the intensity
can be written
\begin{equation}
I(x,y)=2P/\pi w_{0}^{2}~e^{-2x^{2}/w_{0}^{2}}~\chi _{\lbrack
y0-d/2,y0+d/2]}(y),
\label{EqIntensityDistribution}
\end{equation}
where $w_{0}\approx 1~\mathrm{cm}$ is the beam waist and $P\approx 100~%
\mathrm{mW}$ the laser power. The laser light has a well-defined
polarization $\sigma ^{\pm }$ or $\pi $. The atomic beam is
assumed to
provide a homogeneous atomic density in the interaction region $n(\mathbf{r}%
)\approx n_{0}\approx 9\times 10^{6}~\mathrm{cm}^{-3}$.

The fluorescence is collected along the $y$-axis, \emph{i.e.} we
integrate the fluorescence signal across the thin slice. A lens
system images the scattered light so that we obtain a spatially
modulated photon flux which can be expressed through a
generalization of equation~(\ref {EqResonanceFluorescence}) by
\begin{eqnarray}
&&S_{\Delta ,y_{0},\varsigma }(x,z)  \label{EqImage} \\
&=&\alpha t_{probe}\int\nolimits_{-d/2}^{d/2}\frac{I(x,y)}{\hbar \omega }~n(%
\mathbf{r})\sigma (\mathbf{r},\Delta )dy  \nonumber \\
&\approx &\frac{\alpha t_{probe}~n_{0}d\Gamma \Omega _{\mathbf{\hat{%
\varepsilon}\hat{B}}}(x,y_{0},z)^{2}}{4\left[ \Delta -\hbar
^{-1}\varsigma
\mu _{B}~|\mathbf{B}(x,y_{0},z)|\right] ^{2}+2\Omega _{\mathbf{\hat{%
\varepsilon}\hat{B}}}(x,y_{0},z)^{2}+\Gamma ^{2}}.  \nonumber
\end{eqnarray}
The $z$ dependence of the Rabi frequency comes from the
dependence of the coupling strength on the relative orientation
of $\mathbf{\hat{\varepsilon}}$ and $\mathbf{B}$. The photon flux
arriving at the image plane is recorded with a CCD camera. The
proportionality constant $\alpha $ can be estimated knowing the
camera pixel size $d_{pxl}=9~\mathrm{\mu m}$, its sensitivity
(including quantum efficiency and electronic gain) $\eta _{CCD}=1~\mathrm{%
count}/100~\mathrm{photons}$, the light collection solid angle of
the imaging lens system $\Omega _{sa}=0.06\%$, and its
magnification $\varsigma _{mag}=0.17$,
\begin{equation}
\alpha =\frac{d_{pxl}^{2}}{\varsigma _{mag}^{2}}~\eta
_{CCD}\Omega _{sa}.
\label{EqConstant}
\end{equation}
We typically integrate the signal for $t_{probe}=500~\mathrm{ms}$.

Equation~(\ref{EqImage}) immediately reveals that, because we
assumed strong magnetic fields, the fluorescence nearly vanishes
everywhere where the resonance condition $\Delta =\pm \hbar
^{-1}\mu _{B}|\mathbf{B}(\mathbf{r})|$ is not satisfied, or in
other words, the laser light slices the magnetic field at a given
field strength $|\mathbf{B}(\mathbf{r})|=\mp \hbar \Delta /\mu
_{B}$. The spatial distribution of the fluorescence reflects the
\textit{modulus} of the magnetic field. This situation is
depicted in Figure~2 by a plane that interpenetrates the magnetic
potential. For a given detuning $\Delta $, light sheet position
$y_{0}$, and polarization $\lambda $ we expect the fluorescence
image $S_{\Delta ,y_{0},\varsigma }(x,z)$ to form a narrow ridge
along a closed path which basically traces a two-dimensional
equipotential contour line of the magnetic field constraint to
the plane of the laser light sheet (see Fig.~5). At the top of
the fluorescence ridge the photon count rate reads
\begin{eqnarray}
S_{\Delta ,y_{0},\varsigma }(x,z) &\approx &\alpha t_{probe}~\frac{%
n_{0}d\Gamma \Omega _{\mathbf{\hat{\varepsilon}\hat{B}}}(x,y_{0},z)^{2}}{%
2\Omega
_{\mathbf{\hat{\varepsilon}\hat{B}}}(x,y_{0},z)^{2}+\Gamma ^{2}},
\label{EqRidge}
\end{eqnarray}
where $\hbar \Delta =\varsigma \mu _{B}B(\mathbf{r}).$ We see
that the location of the ridge is not influenced by the
dependence of the Rabi frequency on
$\mathbf{\hat{\varepsilon}\hat{B}}$, but the height of the ridge
is modulated. Furthermore, the contrast of this modulation
vanishes as saturation is approached, $\Omega \rightarrow \Gamma
$. Another reason for a modulated fluorescence is the imperfect
homogeneity of the atomic density $n(\mathbf{r})$. This
contribution, however, is easily avoided by normalizing the
fluorescence image with an image at zero magnetic field.

As we stated earlier, the cloverleaf trap is, in axial direction,
a magnetic bottle, \emph{i.e.} the trapping potential is situated
on top of a large
magnetic field offset pointing in $z$-direction. In this case, $B_{z}/|%
\mathbf{B}|\approx +1$, and we get the whole equipotential line
using a single $\sigma ^{+}$ (or $\sigma ^{-}$) laser
polarization. However, this does not hold true for very
compressed traps when the field offset is compensated by a
homogeneous antibias field. Then in some regions of space, we may
have $B_{z}/|\mathbf{B}|<0$. For the same reason, quadrupolar
fields
(or more generally, fields which are antisymmetric with respect to the $z$%
-axis) are only half imaged with a single circularly polarized
laser. However, if we irradiate linearly polarized light, which
is a linear superposition of left and right circularly polarized
light, we get the complete image $\sum\nolimits_{\varsigma =\pm
1}S_{\Delta ,y_{0},\varsigma }(x,z)$.

\section{The Experiment}
\label{sec:3}

In our experiment, we use a fast atomic beam to probe the
magnetic field. Hot sodium atoms are ejected from a heated oven
through a fine nozzle. After
a $2~\mathrm{m}$ long distance, they arrive at the interaction region as a $%
2~\mathrm{cm}$ wide highly collimated and homogeneous beam: We
verified that the radial velocity distribution $v_{\bot }$ is
narrow enough to allow us to
neglect the transversal Doppler effect at the $D2$ atomic resonance, $%
kv_{\bot }<\Gamma $. The atoms enter the magnetic field with a
velocity satisfying $\partial _{t}B/B\ll \Omega _{L}$ which allows
the assumption that the atomic spin adiabatically follows the
fields.
\begin{figure}
\vspace{10.5cm}
\caption{(a) Calculated and measured fluorescence
image of a light
scattering atomic beam inside a cloverleaf-shaped magnetic field with $%
I_{clov}=285~\mathrm{A}$ and $I_{anti}=80~\mathrm{A}$. The laser
is linearly polarized and detuned $+50~\mathrm{MHz}$ from
resonance. (b-d) Equipotential lines derived from fluorescence
images as shown in Fig.~(a). The total current for the magnetic
trap is $I_{clov}=285~\mathrm{A}$. In Fig.~(b) the
antibias current is $I_{anti}=85~\mathrm{A}$, the detuning is $+50~\mathrm{%
MHz}$, and the light is $\sigma ^{-}$ polarized. The different
lines correspond to different locations of the laser slice
displaced from one another by $1~\mathrm{mm}$. In Fig.~(c) the
slice is kept at the center, the
antibias current at $I_{anti}=80~{A}$, and the detuning is ramped from $+10~%
\mathrm{MHz}$ to $+50~\mathrm{MHz}$. In Fig.~(d) the slice is in
the center, the detuning is fixed at $+25~\mathrm{MHz}$, and the
antibias current is ramped from $I_{anti}=80~\mathrm{A}$ to
$I_{anti}=100~\mathrm{A}$.}
\label{fig:5}
\end{figure}

\bigskip

In the interaction region the atoms cross a beam of near-resonant
laser light. When magnetic fields are present, the fluorescence
scattering is spatially modulated by the inhomogeneous
Zeeman-shift of the resonant levels. The fluorescence can
conveniently be imaged to a CCD camera. Blurring of the images by
column integration along the imaging direction can be avoided by
only irradiating a thin sheet of light whose plane is orthogonal
to the imaging axis. In this way, two spatial dimensions of the
magnetic field are directly imaged and the third is probed by
moving the light sheet back and forth. The strength of the
magnetic field is measured by tuning the laser light to the
Zeeman-shifted resonances, thus slicing the magnetic potential at
a depth that can precisely be set by the laser frequency (see
Fig.~2).

Our experiment is sketched in Figure~4, and an example of a
fluorescence image in shown in Figure~5(a). For this image we
operated our cloverleaf trap at full current
$I_{clov}=285~\mathrm{A}$, applied $I_{a}=80~\mathrm{A}$ antibias
current, and located the laser light sheet at the center of the
trap. The ring-shaped border corresponds to atoms being resonant
with laser light blue-shifted by $\Delta =2\pi \times
50~\mathrm{MHz}$. Therefore, the magnetic field along this border
line is $|\mathbf{B}(\mathbf{r})|=\hbar \Delta /\mu
_{B}=36~\mathrm{G}$.
\begin{figure}
\vspace{8.8cm}
\caption{(a) Multiple magnetic field tomographic
lines. When the dynamic range of the magnetic field strength is
large enough, we observe the Zeeman splitting of the $\sigma
^{+}$ transitions as multiple tomographic
lines. The figures correspond to the cloverleaf trap parameters $%
I_{clov}=285~\mathrm{A}$ and $I_{anti}=60~\mathrm{A}$, and the detuning is $%
+20~\mathrm{MHz}$. The four ridges correspond from the inside to
the outside
to the transitions $(m_{F},m_{F}^{\prime })=(2,3),$ $(1,2),$ $(0,1),$ and $%
(-1,0)$. The noise in the measured date has been reduced to
increase the visibility of the features. (b) and (c) The
fluorescence light is polarized, and the polarization depends on
the local magnetic field. In the figures (b) it has been filtered
with a polarizer oriented to transmit light being linearly
polarized along $\mathbf{\hat{\varepsilon}}=\mathbf{\hat{e}}_{z}$.
In the figures (c)
$\mathbf{\hat{\varepsilon}}=\mathbf{\hat{e}}_{x}$ light is left
through.}
\label{fig:6}
\end{figure}

By recording images of the equipotential lines for various
detunings and various positions of the laser light slices along
the imaging axis, we are able to reconstruct the magnetic field
in three dimensions. To demonstrate this, we record several
images like the one shown in Figure~5(a) with varied parameters.
From the fluorescence maximum of every recorded image, we extract
the closed ridge. For this purpose, we have written a data
analyzing program which, from an arbitrary starting point, is
capable of climbing the closest maximum and discerning the path
to follow the ridge. We plot all ridges into a common picture.
For example, the right hand side of
Figure~5(b) shows the equipotential lines for $\Delta =2\pi \times 50~%
\mathrm{MHz}$ and $I_{anti}=40~\mathrm{A}$ at various positions
of the
slice, thus forming an equipotential surface, $|\mathbf{B}(\mathbf{r}%
)|=const $. If we know the surfaces for all values of $const$, we
can
reconstruct the three-dimensional magnetic vector field $\mathbf{B}(\mathbf{r%
})$. To record equipotential surfaces at other magnetic field
strengths, we repeat the procedure for different laser detunings.
This is demonstrated in Figure~5(c) where we kept the light sheet
at the trap center and varied the detunings between
$10~\mathrm{MHz}$ and $50~\mathrm{MHz}$. Figure~5(d) shows
the equipotential lines for fixed detuning of $\Delta =2\pi \times 50~%
\mathrm{MHz}$ and centered light sheet position but varied
antibias current.

The left hand sides of the Figures~5(a-d) show the result of
calculations using formula~(1). In order to improve the
calculations, we calibrated them by measurements using a
procedure to be detailed in the following paragraph. The
remaining small discrepancies between theory and experiment are
probably due to the inhomogeneous intensity distribution within
the laser beam.

\bigskip

When we drive the $D2$ line with $\sigma ^{+}$ light, five
transitions
between the hyperfine levels $^{2}S_{1/2},$ $F=2$ and $^{2}P_{3/2},$ $%
F^{^{\prime }}=3$ satisfy the selection rule $m_{F^{\prime
}}-m_{F}=1$. Under certain conditions, we experimentally observe
all of them simultaneously, since the laser can be resonant with
different transitions simultaneously at different regions of
space, where the magnetic field has different values
$|\mathbf{B}(\mathbf{r})|=\hbar \Delta /\left[ \mu
_{B}(g_{F^{\prime }}m_{F^{\prime }}-g_{F}m_{F})\right] $. The
right hand side of Figure~6(a) shows such a fluorescence image.
When performing the calculations (left hand side of the figure),
it is essential to take into account the nonlinear dependence of
the Zeeman-shift on the magnetic field for transitions between
non-stretched states (Paschen-Back regime of the hyperfine
structure). For calculating the curves in Fig.$~$6(a), we
numerically diagonalize the matrix~(\ref{EqIntermediateShift})
for the ground and excited hyperfine level.

Furthermore, we observe that the fluorescence light is polarized
and that the polarization of is spatially modulated by the
magnetic field. For the images shown in Figure~6(b) and (c), we
filtered the fluorescence light in front of the camera with a
linear polarizer. The fluorescence ridge does not move, however,
its contrast is strongly modulated and changes when we set the
transmitting axis of the polarizer along
$\mathbf{\hat{\pi}=\hat{e}}_{z}$ (Figure~6(b)) or
$\mathbf{\hat{\pi}=\hat{e}}_{x}$ (Figure~6(c)). The behavior is
easily understood in terms of the radiation pattern of an atomic
dipole inside a magnetic field. In all directions orthogonal to
$\mathbf{B}$ the atom emits linearly polarized $\sigma$ light. In
the plane which
is sliced by our laser light sheet, our cloverleaf trap yields $\mathbf{B}%
(x,0,z)\perp \mathbf{\hat{e}}_{y}$, so that we expect linearly
polarized light, indeed. And since $\mathbf{B}(x,0,z)$ rotates
along the fluorescence ridge, the polarization of the scattered
light does so, as well. The left two pictures show the result of
calculations according to
\begin{equation}
S_{\Delta ,y_{0},\varsigma }^{(\pi )}(x,z)=S_{\Delta ,y_{0},\varsigma }(x,z)%
 \frac{|\mathbf{B(\mathbf{r})\times \hat{\pi}|}}{|\mathbf{B(\mathbf{r}%
)|}}.  \label{EqPolarized}
\end{equation}

\bigskip

The resolution of the MFT method can be defined as the halfwidth
of the fluorescence ridge observed at the CCD camera. With this
definition, the width $2~\delta z$ of the emerging contour line
in $\mathbf{\hat{e}}_{z}$ direction immediately follows from
equation~(\ref{EqImage}), $2\delta
z=\hbar \sqrt{\Gamma ^{2}+2\Omega _{\mathbf{\hat{\varepsilon}\hat{B}}}^{2}}%
/(\mu _{B}\partial _{z}|\mathbf{B}|)$. Thus it depends on the
magnetic field gradient and on the (power-broadened) linewidth of
the $D2$ resonance. In
terms of magnetic fields the resolution is given by $\delta B=\partial _{z}|%
\mathbf{B}|~\delta z$, and we see that, even if we work below saturation, $%
\Omega \ll \Gamma $, we are not sensitive to magnetic field
variations that are weaker than $\delta B\approx 3.6~\mathrm{G}$.
Figure~7 shows cuts through fluorescence images along a line
defined by $x=y=0$ for various laser intensities. The
fluorescence rate increases proportionally with low laser
intensities, but saturates at high intensities. The highest
spatial resolution found was about $1~\mathrm{mm}^3$ at a location
where the axial field gradient\ was $\partial
_{z}|\mathbf{B}|\approx 50~\mathrm{G/cm}$ which corresponds to
$\delta B\approx 5~\mathrm{G}$. We could, in principle, use
narrower (Raman) transitions. However, the fluorescence rate will
be lowered as well, thus requiring a more sensitive detection
system.
\begin{figure}
\vspace{7cm}
\caption{Cuts through fluorescence images along a line defined by $%
x=y=0$ for various laser intensities $I=5.6,12,118,$and
$186~\mathrm{mW}$. The effect of power-broadening is clearly
visible. The magnetic field
parameters were $I_{clov}=285~\mathrm{A}$, $I_{anti}=85~\mathrm{A}$ and $+40~%
\mathrm{MHz}$.}
\label{fig:7}
\end{figure}

\bigskip

The MFT method can also be used to precisely determine the amount
of antibias current necessary for exact compensation of the
offset field created by the pinch coils. For this purpose, we set
the antibias current to some value $I_{anti}$, observe the
fluorescence distribution (Fig.~5(a)) on-line, and find the laser
detuning $\Delta $ at which the fluorescence disappears. This
gives us the minimum of $B_{z}$. We repeat this procedure for
various antibias currents and obtain two curves, one for $\sigma
^{-}$
and one for $\sigma ^{+}$ polarized light. The curves intersects near $%
\Delta =0$. At this detuning the antibias is perfectly compensated, \emph{%
i.e.} the magnetic trap has maximum compression. The curves are
shown in Figure~8. The compensation current is $I_{anti}=(86.1\pm
1.1)~\mathrm{A}$. As we see, the uncertainty of this method is
rather small. Furthermore, the
slopes of the curves in Figure~8 yield the calibration of the pinch coils, $%
B_{pnch}/I_{clov}=0.830~\mathrm{G/A}$, and of the antibias coils, $%
B_{anti}/I_{anti}=2.743~\mathrm{G/A}$, at the center of the
trap.\ We can now utilize these data to correct our magnetic
field calculations based on Formula~(\ref{EqBiotSavart}). We also
calibrated the current $I_{clov}$ using the same method and
obtained values for the cloverleaf trap parameters that agree
well with the above ones.
\begin{figure}
\vspace{7cm}
\caption{Compensation of the magnetic field offset via $I_{anti}$
using $I_{clov}=285~\mathrm{A}$. Shown is the detuning at which
atoms at the bottom of the potential scatter light. These data
can be used to calibrate the calculations of the magnetic
fields.} \label{fig:8}
\end{figure}

\section{Conclusion}
\label{sec:4}

In conclusion, magnetic fields have been probed using a technique
well-suited for fast three-dimensional tomographic reconstruction
of a magnetic field within a volume which is not accessible to
material detectors like Hall probes, \emph{e.g.} within an
ultra-high vacuum. The drawback of this method is that the
resolution is obviously limited by the natural linewidth of the
transition, so that it is only applicable for large magnetic
fields and large field gradients. A solution to this problem
could be the use of radiofrequency-optical double resonance
schemes \cite {Martin88,Helmerson92}.

The MFT method is well adapted to investigating magnetic trapping
fields for neutral atoms. In particular, it permits precise
mapping of the trap geometry and fast localisation of its minimum.
The response time of the MFT method is only limited by the time
required for optical pumping. Therefore the method could be
extended for monitoring rapid changes in the magnetic field
strength being as fast as a few $10~\mathrm{ns} $ using
stroboscopic imaging or even for real-time control of the magnetic
trap configuration. We conclude that, since this technique only
requires a hot atomic beam and a weak tunable laser beam, we
believe it to be a simple and versatile diagnostic instrument for
probing magnetic fields.

\bigskip

\begin{acknowledgement}We acknowledge financial support from the
S\~{a}o Paulo State Research Foundation FAPESP and Ph.W.~C.
wishes to thank the Deutscher Akademischer Austauschdienst DAAD
for financial support. R.K. also acknowledges support from the
Centre National de la Recherche Scientifique CNRS.
\end{acknowledgement}

\end{document}